 \documentclass[final,5p,times,twocolumn]{elsarticle}

\usepackage{amssymb}
\usepackage{lipsum}

\journal{Nuclear Physics B}

\begin{document}

\begin{frontmatter}



\title{Reminiscences about Steven Weinberg (This Time it's Personal)
}


\author[first,second,third]{C.P. Burgess}
\affiliation[first]{organization={Department of Physics \& Astronomy, McMaster University},
            addressline={1280 Main Street West}, 
            city={Hamilton},
            postcode={L8S 4M1}, 
            state={ON},
            country={Canada}}
\affiliation[second]{organization={Perimeter Institute for Theoretical Physics},
            addressline={31 Caroline Street North}, 
            city={Waterloo},
            postcode={N2L 2Y5}, 
            state={ON},
            country={Canada},}
\affiliation[third]{organization={School of Theoretical Physics, Dublin Institute for Advanced Studies},
           addressline={10 Burlington Rd.}, 
           city={Dublin},
           state={Co. Dublin}, 
           country={Ireland},}

\begin{abstract}
Steven Weinberg productive scientific life teaches us many things, one of the most important of which is the power of his example. This essay contains personal reminiscences and a speculation about gravitational wave propagation based on one of his very last papers -- as seems fitting, given his penchant for putting interesting physics into the essays he wrote for other luminaries over the years. (See \cite{Burgess:2025fqp} for a less personal summary of his main scientific accomplishments.)
\end{abstract}







\end{frontmatter}




\section{Reminiscences}
\label{sec:Reminiscences}

I first heard about Steven Weinberg when I stumbled across, and then read, his book {\it The First Three Minutes} \cite{Weinberg:1977ji} sometime as an undergraduate. But I don't think he made an impression on me as a scientist (as opposed to a writer) until the fall of 1979, when he won his Nobel prize for electroweak unification. It was my last year of undergraduate studies and I had just learned about the `gauge principle'. So that Nobel prize caught my attention and confirmed my earlier instincts to study fundamental physics if I could. I was also keenly aware of how much there still was to learn, even after completing an entire undergraduate degree in physics. 

I didn't think much more about it until a phone call later that year changed my plans for future study. I had applied to the usual suite of graduate schools and was accepted into about half of those to which I'd applied. I had attended the conference for physics undergraduates that is organized annually by the Canadian Association of Physicists and had been very impressed there by John Wheeler, who was one of their keynote speakers, and so also applied to the University of Texas (to which Wheeler had moved relatively recently). One night early in the winter term I received a phone call out of the blue and was gobsmacked to discover it was Wheeler on the line, offering a fellowship to come to Texas. He said Steven Weinberg was about to move there as well and that news tipped me over the decision point. 

In the summer before graduate school I read Weinberg's textbook {\it Gravitation and Cosmology} \cite{Weinberg:1972kfs} in some detail, and it made a great impression on me. I had learned General Relativity (GR) in undergraduate courses, but these were taught in the Applied Mathematics department and it all seemed very abstract. GR in Weinberg's telling was much more visceral: it was a book full of the physics of things falling. It was only at that point that I thought `I can see myself working on this subject'. 

\begin{figure}
	\centering 
	\includegraphics[width=0.4\textwidth, 
	]{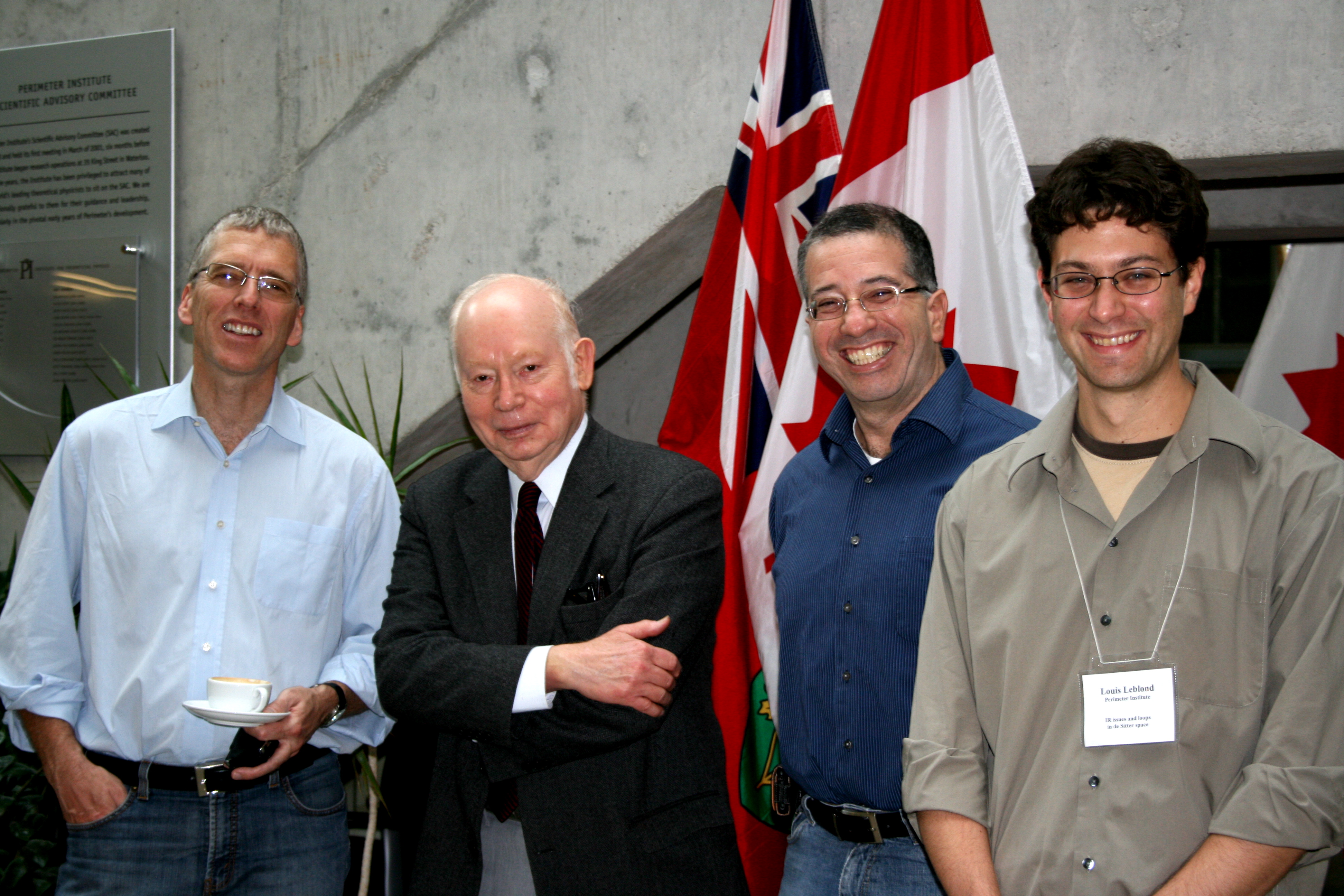}	
	\caption{The author, Steven Weinberg, Richard Holman and Louis Leblond while attending {\it de Sitter Days} at Perimeter Institute} 
	\label{fig_mom0}%
\end{figure}

\subsection{Graduate school}

I arrived to Austin to start my PhD in a swelteringly hot day in August 1980. Because of my fellowship I was given an office on the 9th floor - prime real estate directly across the hall from John Wheeler and a few doors down from Weinberg. The first time I saw Weinberg in the flesh I wondered where the Secret Service agents were -- it seemed odd to find him wandering around the department just like everyone else. 

Although I did not appreciate it at the time, in retrospect Texas provided a unique environment in which Particle Physics, General Relativity and Quantum Gravity were actively explored cheek by jowl and it was possible for a student to move seamlessly amongst them. Particle physicists (largely driven by Weinberg) at the time were discovering how gravity (in particular supergravity) could play an important role in answering questions about the electroweak hierarchy. Courses by Bryce DeWitt, Philip Candelas, Joe Polchinski, Willy Fischler, Claudio Teitelboim, Duane Dicus, George Sudarshan, Ilya Prigogine and others gave the impression that physics was an open range relatively unencumbered by fences dividing it into narrow paddocks of specialization. 

But best of all were Weinberg's courses, first on Supersymmetry and later on Quantum Field Theory (collectively the subject matter of his three-volume textbooks on Quantum Field Theory). He would hand out xeroxed copies of hand-written notes which he would then never consult during his lectures while performing calculations in real time on the blackboard. Notes would only be referred to occasionally at the very end to verify his answer was correct. Almost as useful as the material itself was seeing his thought processes when computing: calculational tricks for quickly arriving at the answer and error-correcting processes `that cannot be right because..'. One came out of those courses with an appreciation for his famously bulldozer-like style: a very broad first-principles approach to writing down the most general way to handle physically well-motivated problems. First principles were important because they show which features of a theory can be tinkered with and which are more sacred. 

On my arrival Philip Candelas took me under his wing, both by inviting me on one of his fabled rafting expeditions on the Rio Grande in Big Bend park and by supervising my qualifying exam. The qualifying exam involved giving a seminar to the department, and mine described how to extend the supersymmetry tools from Weinberg's course to describe 2+1 dimensional theories. The usual practice was for the audience to leave after the seminar so the committee could grill the PhD candidate. Weinberg was on my committee, and mine must have been the first one on which he participated in Texas, since he did not seem to know the drill. Once the audience filed out he said ``Well I think he passed, what do you guys think?'' and (seeing it was Weinberg) everyone else on the committee agreed, sparing me the grilling. 

After the qualifier Weinberg invited me to be his PhD candidate and I jumped at the chance. He brought to Texas the group-meeting culture from Harvard and everyone in his group would gather weekly in his office for a brown-bag lunch as someone described their work on the blackboard. Everyone was expected to take their turn presenting, including the students.  In a particularly memorable one of these Weinberg said he'd had a call in which he was told that the W and Z bosons had been discovered at CERN. It was useful to see his mind in action: often after an impenetrable talk Weinberg would say ``so what you're saying is...'' and then concisely summarize the speaker's point. Then he'd ask a constructive question or give some insight. Although he and the rest of the room were a daunting audience for a new student, they were not unfriendly. The example set by the senior people was one in which the important thing was trying to understand the science, rather than grandstanding. 

At one point early on he asked me to stop by his office so we could talk about my thesis topic. I told him maybe it would be better for me to try to find a topic of my own (which he agreed would be a good idea). It later transpired that he then went across the hall and recruited Joe Lykken to help him do the calculations needed to see how supergravity could be the mediator of supersymmetry breaking, culminating in his classic article \cite{Hall:1983iz} with Joe and Lawrence Hall. I learned a valuable lesson the hard way: never turn down an offer from Big Steve\footnote{It was around then that I made the switch from calling him Weinberg to calling him Steve (though he was Big Steve to all the students when discussing amongst ourselves). It took a while for this to feel normal.} to discuss your thesis topic. 

As a student I used to attend colloquia with a view to seeing which areas of physics I should make it a priority to learn. Colloquia are usually good ways to learn the Big Picture of a subject because they are given by experts but the information is presented at a level that a student can often understand. In one of these John Schwarz described progress in String Theory as a quantum theory of gravity and based on the skeptical response of the audience I decided it was not something I'd ever have to learn. The Great String Revolution occured the very next year, in which the discovery of anomaly cancellation tipped the majority view in favour of it having a chance to be right. This was one of my first experiences of having to update my scientific priors, and of seeing much of the field also do so.

It was particularly instructive to watch locals like Steve and Joe Polchinski change their minds and start to teach themselves string theory. Weinberg did so in typical bulldozer fashion, setting up the general way that 2-dimensional correlation functions could be interpreted as higher-dimensional scattering amplitudes. Polchinski, by contrast, did so by working out one-loop string scattering on a torus. It was beautiful to watch: he found the simplest calculation that allowed access to the key ideas. Both approaches were instructive and contrasting them spoke volumes about the different styles taken by people with different taste as they approach the same physics.  
 
Steve's main gift to me as a graduate student was his grand vision of physics and how Effective Field Theories (EFTs) capture so much of it so efficiently. This was part of the {\it zeitgeist} during the brown-bag lunches and also came across in his Quantum Field Theory classes. When the power of the message sank in it was for me one of those burning-bush epiphanies that happen only a few times in one's life. As a graduate student I took the EFT point of view for granted, as one does. I did not appreciate that it wasn't equally well understood elsewhere, and this was unfortunate because working it out in detail -- as was eventually done by others in the literature -- would have provided a number of very good thesis topics. In later life when I wrote a book on the subject \cite{Burgess:2020tbq} he said although he had a policy never to write endorsements for book jackets he would be very happy to provide a review of the book once it was out. Sadly he passed on before being able to do so.

\subsection{Later Interactions}

Weinberg's influence remained in the background after I graduated and grew up scientifically. On graduating I moved from Texas to a postdoctoral position at the Institute for Advanced Study (IAS) in Princeton, followed by a faculty position at McGill University in Montreal. Almost twenty years later I moved to McMaster University and Perimeter Institute. I have no doubt that his support helped propel me along at the various inflection points of my career. 

Princeton during my postdoctoral tenure was a hotbed of String Theory and every month seemed to bring a new surprise. It was a perfect place for a newbie to learn the subject because with all those experts around it was simple to differentiate between the things only I didn't know from the things nobody yet knew. It was shortly after writing a paper (together with fellow postdoc Tim Morris) on how to formulate string calculations {\it \`a la Polyakov} in open and unoriented string theories that I received my first phone call from Steve, asking probing questions about world-sheet duality (and helpfully pointing out errors I had made). Such calls could be a shock particularly when they came early in the morning.   

While at McGill I had the great pleasure of successfully nominating Steve for an honorary degree. He gave the convocation address on the day of the event and said that because it was the first time he had spoken at a convocation involving only science students (as opposed to scientists and engineers together) he intended to take the opportunity to give advice to young scientists. This advice was later published as an essay \cite{GoldenLessons} and it is worth reading, even by mature scientists. 

It was around this time that he was developing a taste for controversial public debate and one of his convocation messages was about how science had undermined the ability of religion to be taken as a face-value explanation of the workings of the natural world. As this message emerged in his address you could hear the audience -- largely the graduates' family members and friends -- getting quieter and quieter, but it was artfully done and very well received. Once he sat down the ceremony turned to the next item on the agenda: a cleric rose to give the benediction, perfectly breaking the ice.  

Over the years we would interact from time to time, largely by email. It was during a visit to Texas to give a seminar in the early 90s that I learned I had sent Steve his very first email message. When introducing me for the talk he said that every time he opened his mailer (PINE) my name was the first thing he would see. My name came up because he left all messages in the inbox and they were displayed in chronological order in which they'd been received. 

I was his host when he came to Perimeter Institute as a keynote speaker for the first graduation ceremony for the Perimeter Scholars International (PSI) program of Master's students, and he returned several times to speak about cosmology and asymptotic safety. In all of these talks he described his recent research; never falling into the Wise Old Man role that would have been so easy to take. 

Indeed, he was supremely successful in evading the trap he felt other luminaries such as Einstein had failed to avoid: he never allowed his spectacular successes to stop his doing journeyman calculations that others might have felt to be beneath the ambition of a Nobel laureate. In this way his physics voice remained authentic and personal right up to the end.

\section{A bit of physics}
I would like to close this essay with a physics speculation based on one of his very last papers \cite{Flauger:2019cam} that was written not long before his death. I do so partly to celebrate the great longevity of his more than 60 years of active research.

But I also do so to salute his knack for inserting novel physics ideas into his own testimonial essays for other physics luminaries. Two good examples of this are {\it Why Renormalization is a Good Thing}, which appeared in his essay honouring Francis Low \cite{Weinberg:1981qq}, and {\it Superconductivity for Particular Physicists} which appeared in his essay for Yoichiro Nambu \cite{Weinberg:1986cq}. He similarly introduced the notion of Asymptotic Safety in his Erice lectures on Critical Phenomena \cite{Weinberg:1976xy}. 

In their paper Weinberg and Rafael Flauger compute the absorption rate of gravitational waves by inverse bremstrahlung in the collisions of electrons and ions in hot interstellar gas. Gravitational waves had only recently been observed on Earth by the LIGO collaboration and it is a testament to their intellectual agility that they immediately began to think about calculations that could help illuminate what we might learn from their observation. It is a beautiful paper that draws on Steve's own seminal calculations of soft graviton emission and absorption by hot particles \cite{Weinberg:1965nx} and so in a sense completes a circle in his intellectual trajectory. 

They compute the absorption rate in the limit\footnote{These expressions use fundamental units for which $\hbar = c = k_{\scriptscriptstyle B} = 1$.} $\nu \ll  T$ where $\nu$ is the gravitational wave's frequency and $T$ is the temperature of the hot gas, and find that it is given by the expression 
\begin{equation}
  \Gamma_{\rm abs} \simeq \frac{G m_e^2 n_e^2 }{5\pi^2 \nu^3} \left\langle u^5 \sigma \right\rangle \,.
\end{equation}
Here $G$ is Newton's constant of gravitation, $m_e$ is the electron mass, $n_e$ is the electron density, $u$ is the electron speed and $\sigma$ is the scattering cross section for the reaction ({\it e.g.}~Coulomb scattering) that causes the electrons to accelerate. The angle brackets $\langle \cdots \rangle$ denote any relevant average (such as thermal) over the initial velocities. 

My first reaction to this paper at the time was `why do this?' because my intuition -- and that of most others -- was that gravitational interactions are so weak that there will never be interesting effects. Weinberg and Flauger nonetheless press on to observe that the strong $\nu^{-3}$ dependence for small $\nu$ means that this result can be surprisingly large. For Coulomb scattering amongst hot electrons with temperature $T \sim \hbox{keV}$ and densities $n_e \sim 10^{-3}$ cm${}^{-3}$ it predicts an attenuation length of order 1 Mpc for frequencies smaller than 240 nHz (reasonable numbers for interstellar gas in galactic clusters and gravitational waves observable through measurements of pulsar timing). If this were the full story the Universe would be too opaque to gravitons to allow them to be seen from sources as distant as those from which higher frequency gravitational waves have already been seen. 

Of course Steve and Rafael also show that the above formula is {\it not} the whole story; this absorption rate is partially cancelled by the rate for stimulated emission that also varies like $\nu^{-3}$ for small $\nu$, leading to a suppression of the {\it net} absorption by a factor of $2\pi \nu/T \sim 10^{-24}$. The visible Universe is transparent to gravitational waves, but only because of this cancellation. Should pulsar timing reveal gravitational waves from cosmologically distant sources then we would know that these waves must experience stimulated emission within the environment through which they move.

Although the final result is an effect that is not (yet) observable, it is also a classic illustration of the utility of explicit calculation even if the results are unlikely to be immediately detectable. In his own words in another context Weinberg put it this way \cite{Weinberg:2005vy}: 
\begin{quote}
   {\it ...just as field theorists in the 1940s and 1950s took pains to understand quantum
electrodynamics to all orders in perturbation theory, even though it was only possible to verify results in the first few orders.}
\end{quote}
But even more remarkable (at least to me) is the observation that the above cancellation shows that gravitational waves can be in a regime where interactions with their environment are significant, contrary to common intuition. 

I find this interesting because experience with electromagnetic waves teaches us that this kind of regime sometimes opens the door to unexpectedly large medium-dependent effects in wave propagation. After all, electromagnetic waves also interact relatively weakly with matter due to the small size of the electromagnetic coupling constant $\alpha = e^2/4\pi \simeq 7 \times 10^{-3}$, where $e$ is the elementary charge. Yet medium-dependent effects (like the in-medium propagation speed $c$ of light) can differ from their vacuum counterparts by order unity (as is the case for light propagation in water or glass). We do {\it not} find $c = 1 - {\cal O}(\alpha)$ in these cases even though all electromagnetic couplings to matter are explicitly proportional to $e$.  

The origin of such large effects in electromagnetism is well understood, and is a consequence of an interesting cancellation.\footnote{The discussion here follows the one given in \S{16.3.3} of \cite{Burgess:2020tbq}.} For polarizable media the size of the medium-induced change to the propagation speed is controlled by the size of the product $\kappa\,  n$, where $n$ is the density of atoms and $\kappa$ is their polarizability. In simple atoms the polarizability is computed by calculating the atom's leading ${\cal O}(E^2)$ energy shift in the presence of an electric field of magnitude $E$. This arises at second order in perturbation theory and has the schematic form
\begin{equation}
    \kappa \sim \frac{\vec{D} \cdot \vec{D}}{\Delta E} \,,
\end{equation}
where $\vec{D} \propto \langle n' | e \vec{x} |n \rangle$ is the transition matrix element between electron states of the dipole moment operator and $\Delta E$ is the energy difference between these states. 

Now comes the main point: although $\vec D$ indeed is proportional to the electromagnetic coupling -- with magnitude $D \sim e a_{\scriptscriptstyle B}$ where $a_{\scriptscriptstyle B}$ is the size of the atom --  this cancels in $\kappa$ because $\Delta E \propto e^2/a_{\scriptscriptstyle B}$, leaving an $e$-independent result that is order $\kappa \sim a_{\scriptscriptstyle B}^3$. Using this in the product $\kappa \, n$ shows that $\kappa\, n \ll 1$ when atoms are not closely packed relative to their size (like in air). For such materials one expects medium-dependent effects to be small: $c -1 \sim  {\cal O}(\kappa\, n) \ll 1$. But crucially these effects need not be small when atoms {\it are} closely packed (such as in water or glass) since then $\kappa \,n$ and $c-1$ can both be order unity. The weakness of the electromagnetic coupling only enters into the size of $\kappa$ indirectly, through its role in setting the size $a_{\scriptscriptstyle B}$.

This is all very nice, but it only really matters for light propagation in a regime where the amplitude for photon absorption and re-emission is not negligible. In the present context the relation between $\kappa \, n$ and the propagation speed of a light wave is derived in the mean-field approximation in which the medium's fluctuations about the mean are small. If fluctuations are not small they tend to scatter photons and make the medium opaque, so for interesting effects one seeks situations where propagation effects like $c - 1$ are order unity but `diffuse' scattering is negligible. For photons this regime relies on the wavelength of the photon being much larger than the distance between scatterers and on the coherence of photon scattering in the forward direction. 

My prejudice before reading the Flauger-Weinberg paper was that the incredibly weak interaction strength of gravity would make the criterion of significant absorption and re-emission never apply for gravitational waves, but now I am not so sure. It remains to be seen whether or not environments exist for which a similar cancellation happens for gravity, making medium-dependent effects for gravitational waves surprisingly large. But I think the Flauger-Weinberg observation that medium-dependent absorption and emission can compete significantly even for gravitational waves makes the search seem worthwhile. Success is not guaranteed, but if found would vindicate of another of Steve's maxims  \cite{Weinberg:1977ji}:
\begin{quote}
{\it  
Our mistake is not that we take our theories too
seriously, but that we do not take them seriously enough.} 
\end{quote}  

Looking back, perhaps Steven Weinberg's greatest influence on me was as an example. His deep yet pragmatic vision of physics certainly shaped my own views, but his influence on the whole field has been so pervasive that much the same can be said for most of us currently in the business. It is his personal taste in problems and his clarity and common sense (use sophisticated mathematics if necessary, but not necessarily sophisticated mathematics) 
that remain with me as a standard to which to aspire, as is the culture of integrity he cultivated in those around him. 

His intellectual footprints point a promising way forward, but the shoes he leaves behind for us are truly difficult to fill.

\section*{Acknowledgements}
I would like to thank the editors of this collection for the invitation to contribute an article and their enormous patience awaiting its delivery. It is a privelege to be part of humanity's Great Quest to understand nature. I am grateful both to my many fellow-travellers -- students, colleagues and collaborators -- for their passion and enthusiasm in pursuing it, to the general public for their interest and to their representatives for funds. My own research is supported by funds from the National Sciences and Engineering Research Council (NSERC) of Canada. Research at Perimeter Institute is supported in part by the Government of Canada through NSERC and by the Province of Ontario through the Ministry of Research and Innovation (MRI).










\end{document}